\documentclass[sigconf]{acmart}

\usepackage{csquotes}
\usepackage{todonotes}
\usepackage{breakcites}

\copyrightyear{2024} 
\acmYear{2024} 
\setcopyright{rightsretained} 
\acmConference[WSESE '24]{International Workshop on Methodological Issues with Empirical Studies in Software Engineering}{April 16, 2024}{Lisbon, Portugal}
\acmBooktitle{International Workshop on Methodological Issues with Empirical Studies in Software Engineering (WSESE '24), April 16, 2024, Lisbon, Portugal}\acmDOI{10.1145/3643664.3648213}
\acmISBN{979-8-4007-0567-0/24/04}

\begin{document}

\title{Apples, Oranges, and Software Engineering: Study Selection Challenges for Secondary Research on Latent Variables}

\author{Marvin Wyrich}
\orcid{0000-0001-8506-3294}
\affiliation{%
  \institution{Saarland University}
  \city{Saarbrücken}
  \country{Germany}
}
\email{wyrich@cs.uni-saarland.de}

\author{Marvin Muñoz Barón}
\orcid{0000-0001-5991-3072}
\affiliation{%
  \institution{University of Stuttgart}
  \city{Stuttgart}
  \country{Germany}
}
\email{marvin.munoz-baron@iste.uni-stuttgart.de}

\author{Justus Bogner}
\orcid{0000-0001-5788-0991}
\affiliation{%
  \institution{Vrije Universiteit Amsterdam}
  \city{Amsterdam}
  \country{The Netherlands}
}
\email{j.bogner@vu.nl}

\renewcommand{\shortauthors}{Wyrich et al.}

\begin{abstract}
    Software engineering (SE) is full of abstract concepts that are crucial for both researchers and practitioners, such as programming experience, team productivity, code comprehension, and system security. Secondary studies aimed at summarizing research on the influences and consequences of such concepts would therefore be of great value.
    
    However, the inability to measure abstract concepts directly poses a challenge for secondary studies: primary studies in SE can operationalize such concepts in many ways. Standardized measurement instruments are rarely available, and even if they are, many researchers do not use them or do not even provide a definition for the studied concept. SE researchers conducting secondary studies therefore have to decide a) which primary studies intended to measure the same construct, and b) how to compare and aggregate vastly different measurements for the same construct.
    
    In this experience report, we discuss the challenge of study selection in SE secondary research on latent variables. We report on two instances where we found it particularly challenging to decide which primary studies should be included for comparison and synthesis, so as not to end up comparing apples with oranges. Our report aims to spark a conversation about developing strategies to address this issue systematically and pave the way for more efficient and rigorous secondary studies in software engineering.
\end{abstract}




\maketitle

\section{Introduction}

\emph{The only valid measurement of code quality: WTFs per minute}. 
While one can easily smile at the comic by Thom Holwerda~\cite{Holwerda:2008:WTFmin}, in which the number of swear words used during a code review decides whether code is good or bad, the comic highlights a real challenge that is as old as the software engineering field itself.
Different people have different ideas about what code quality is and how best to measure it.
This is because code quality is a \emph{construct}, i.e. \enquote{a concept that is not directly measurable but is represented by indicators at the operational level to make it measurable}~\cite{Sjoberg:2022:Construct}.

Software engineering (SE) is full of such constructs.
They can enable everyone involved to communicate efficiently on an abstract level, e.g., about a team being more \emph{productive} than in the previous sprint, a software system being sufficiently \emph{secure}, or certain parts of the code not yet being \emph{understandable} enough.
Since all of these constructs can be relevant in practice, SE research is concerned with them: primary studies investigate how software teams can become more productive, which security vulnerabilities make a system no longer secure, and what has a positive and negative effect on the understandability of code.
Secondary studies then summarize the results of related primary studies to inform research and practice about available evidence---or at least that is how it is theoretically meant to be. 

Secondary research often faces a challenge: abstract concepts can have very different meanings in the minds of different researchers. 
And, to date, primary research in software engineering still rarely defines its investigated concepts~\cite{Wyrich:2022:40Years,Tempero:2016:DefiningCoupling}.
An implicit definition by task and measure prevails, but the task and measure may be different in each primary study.
This makes it difficult for someone conducting a literature review to decide which primary study to include to still ensure a meaningful comparison on the conceptual level.
We will illustrate the issue with two examples from our experience with secondary studies in the field of code comprehension.

\subsection{Case 1: An Attempt at Meta-Analysis}

Many attempts were made to find a metric that automatically measures code understandability. A few years ago, we empirically validated such a metric, i.e., we investigated whether a metric proposed by the industry really measures code understandability as it claims~\cite{Munoz:2020:CogComplexity}. However, we did not conduct our own experiment. Instead, we used data sets from ten existing code comprehensibility studies, in which the comprehensibility of certain code snippets had already been measured with human participants. 

We obtained around 24,000 understandability evaluations of 427 code snippets, far more data than we could have collected in a single study.
We were then able to correlate the understandability evaluations with the automated metric values.
This way, we knew how well the proposed metric correlated with the human-collected understandability evaluations within each study.

In a final step, we planned to statistically summarize the correlation coefficients obtained from each study in a meta-analysis. And here we were faced with a difficult decision. Only at this point did we realize how differently primary studies actually measure code comprehensibility: starting from the time needed to calculate a return value, over the correctness to comprehension questions, all the way to the strength of brain deactivation measured as a proxy for concentration through the usage of an fMRI scanner. Statistically, it would have been possible to combine all this data, but would it have made sense?

At the time, we decided against a holistic analysis and instead divided our research question into five sub-questions, each of which was to investigate the correlation of the metric with a specific way of measuring code comprehensibility. To answer each of the sub-questions, we then conducted a separate meta-analysis and created forest plots. Then, as now, we believe that \enquote{while it is a tempting thought to combine the types of variables rather than dividing each of them into separate research questions, in our case the lack of information on the relationships and validity of these measures prevented us from doing so}~\cite{Munoz:2020:CogComplexity}.

Now, we still needed an answer to whether the metric correlated with code understandability. The somewhat sobering answer is: it depends. The metric correlated with comprehension time and subjective ratings of understandability. It correlated less well with the correctness of comprehension tasks and physiological measures. Although these were interesting results, we were a little dissatisfied with the situation, as splitting the research into five sub-questions is almost guaranteed to produce an \enquote{it depends} result. Would we have been comparing apples with oranges if we had thrown all the studies into one bowl? Our results suggest so, but we would only have certainty if we specifically investigated whether all these metrics measured the same construct.

\subsection{Case 2: A Systematic Mapping Study}

The diversity of study designs that seemed to be present in code comprehension studies made us curious. We therefore conducted a systematic mapping study to take a closer look at the design landscape of 95 source code comprehension experiments published between 1979 and 2019~\cite{Wyrich:2022:40Years}. 
However, identifying these 95 studies in the first place was challenging, which can be explained in part by the results of the mapping study: each study in our dataset was unique in its design, they all used different tasks and metrics to measure a study participant's code comprehension, and there was no homogeneity in the naming of the construct of interest. For example, we also examined studies that did not refer to comprehension but to \emph{readability}, \emph{task difficulty}, or \emph{mental effort}. Notably, hardly any of the 95 studies provided a definition of these constructs.

Thus, we were faced with the challenge of speculating about the intention of the authors of the primary studies based on the methodology of the studies: do they all intend to measure the same construct, although they sometimes call it differently, and they all operationalize it somewhat differently?
We read all the papers carefully and decided whether to include a primary study or not to the best of our knowledge and ability. This was more time-consuming and involved more uncertainty than should have been necessary.\\

\noindent
In both of these cases, we eventually found a way to address the challenge of comparing and synthesizing primary studies on code comprehension.
For example, regarding case 2, we decided not to trust the used concept names by the authors, but instead carefully analyzed the experiment designs to find out if the authors really wanted to measure our studied construct, bottom-up source code comprehension from the perspective of human participants.
This required considerably more time for extraction and discussion, with many instances of us angrily throwing our hands up in frustration, but ultimately, it led to much more consistent results.
Nevertheless, it appeared as if we were fighting symptoms rather than a cause, and we did not see an easy way to define generalizable guidance for other researchers based on these experiences.
Variability in study designs is desirable, as it contributes to a comprehensive understanding of a research question. It is also possible, in principle, to make a statistical comparison despite a diversity of study characteristics. However, a critical concern arises regarding the feasibility of meaningful comparisons across primary studies at a conceptual level. This challenge is particularly pronounced when studies operationalize variables in disparate ways without establishing a clear definition of the investigated constructs.
In the remainder of this paper, we first review related literature on this topic (Section~\ref{sec:relwork}), which then forms the basis for our discussion of ways forward in the context of empirical software engineering research (Section~\ref{sec:discussion}).

\section{Related Work}
\label{sec:relwork}

We begin with an overview of the state of secondary studies in SE, followed by a glimpse beyond the borders of SE, incorporating insights from the fields of psychology and medicine.

\subsection{Secondary Research in SE}



Secondary studies in SE research have seen significant advancement since the first push towards evidence-based software engineering (EBSE) in the early 2000s~\cite{Dawson:2003:Empirical}.
Especially when it comes to systematic reviews and mapping studies, there has been a healthy number of publications~\cite{Cruzes:2011:Research, DaSilva:2011:Six}.
Systematic mapping studies have allowed SE researchers to gain quick overviews through the categorization and high-level aggregation of various topics.
To complement this, in-depth systematic reviews then provide researchers with more profound insights into the available evidence about specific topics, often through qualitative synthesis~\cite{Petersen:2008:Systematic}.
Moreover, efforts have been made to establish concrete standards for conducting secondary studies~\cite{Kitchenham:2015:EvidenceBased, Kitchenham:2022:SEGRESS}.
In addition to extensive methodological knowledge about reviews and mapping studies, these guidelines often also include chapters on quantitative synthesis.
But, when it comes to quantitative secondary studies such as meta-analyses, only very few have been published in SE~\cite{DaSilva:2011:Six, Kitchenham:2010:Systematic}.

So, why are there so few meta-analyses in our field?
One possible explanation could lie in the pool of primary research suited for quantitative synthesis.
In fact, difficulties in conducting meta-analysis in SE have been observed even before the advent of EBSE.
At the turn of the millennium, Miller attempted to conduct a meta-analysis on defect detection experiments.
He found that \enquote{the discipline must embark upon a period of improvement to reduce the variability between replicated experiments}~\cite{Miller:2000:Applying}.
The issues brought up by Miller, such as the stark variability in study design, measurement, and reporting of SE experiments, have since been repeatedly found by others~\cite{Jedlitschka:2004:Evidence, Kampenes:2007:Systematic, Brereton:2007:Lessons, Hosseini:2017:Systematic}.
Miller's investigation ended on a positive note, remarking that the field is still very young and things may improve with maturity.
More than 20 years later, we still struggle with the same problems in our attempts at meta-analysis.

Variability in studies, especially in replications, is a desired trait.
Study subjects, settings, materials should differ to provide robustness of the observed effect~\cite{Kitchenham:2008:Role}.
However, variability is undesired when it comes to fundamental understanding and definition of the measured concepts.
If concepts or properties are only the same in name, but are understood differently by researchers, this will often be reflected in a variability in the method of measurement, experiment tasks, and measurement instruments.
Often, researchers do not attempt to evaluate construct validity and do not question the measures they take, which might lead them to incorrect conclusions~\cite{Ralph:2022:Paving}. 
Further, variability in reporting may be addressed with more rigor in the adherence to reporting guidelines.

\subsection{A Look at Other Fields of Research}

Compared to the broader scientific community, software engineering is still a relatively young research area.
This youth presents an opportunity to draw lessons from the rich history of other disciplines, which have faced similar challenges in the past.
Specifically, the field of psychometrics~\cite{Furr:2021:Psychometrics}, with its deep roots in measuring abstract concepts and latent variables, may offer valuable insights to SE researchers.
In psychology, the relationship between human behavior and understanding its expression through quantifiable measures in tests has been discussed since 1886~\cite{Anunciacao:2018:Overview}.

As the years have passed since then, experimental psychologists have faced many of the same challenges that SE researchers still face today.
The problem of multiple, different measures being used for the same abstract concept is a core challenge described in the literature on experimental psychology~\cite{Anunciacao:2018:Overview}.
There, it is argued that the problem is a necessary consequence of the abstract nature of latent variables, as multiple instruments may measure the same psychological phenomenon.
The conclusion to this problem, however, is not the indiscriminate usage of instruments, but rather that the evaluation of \enquote{the attributes of psychological testing is one of the greatest concerns of psychometrics}~\cite{Anunciacao:2018:Overview}.

As a consequence, experimental psychologists have long applied systematic methods to evaluate the validity of psychometric instruments.
Approaches such as classical test theory (CTT)~\cite{DeVellis:2006:Classical}, item response theory (IRT)~\cite{Embretson:2013:Item}, and confirmatory factor analysis (CFA)~\cite{Brown:2015:Confirmatory} could not only be applied when designing instruments to measure latent variables in SE experiments, but also be taught to aspiring SE researchers, as has been the case in experimental psychology for years~\cite{Cousineau:2005:Rise}. 
In fact, when measuring latent variables such as comprehension, SE researchers assume the role of behavioral scientists, as \enquote{they identify some type of observable behavior that they think represents the particular unobservable psychological attribute, state, or process}~\cite{Furr:2021:Psychometrics}.
One might even go so far as to say that measuring the human aspects of SE or conducting experiments about software development constitutes a psychological experiment, and thus, measuring the latent variables in SE is psychometrics.
Controlled experiments in SE very closely mirror those in experimental psychology, but the maturity and rigor of measurement methodology in the latter field could greatly benefit the former.
There have been attempts to bridge this gap and introduce these methods to SE researchers~\cite{Schmettow:2008:Introducing, Graziotin:2021:Psychometrics}, but, as of today, these discussions remain decidedly niche.

In psychometrics, measurement almost exclusively relates to latent human variables, which are measured through meticulously designed instruments.
Software engineering, on the other hand, additionally involves the measurement of software characteristics that are directly measured, such as the number of lines of code.
This difference, however, should not be seen as an excuse to omit discussions about the validity of software metrics and how they represent quality attributes.
For latent human characteristics, we may directly use or adapt measurement instruments used in psychology, but for software metrics, the driving force behind their development is the SE community itself.
Contrary to the lack of discussions about the validity of latent variables, discussions about the validity of software metrics have a much richer history in SE.
As early as 1996, researchers have argued against the blind acceptance of traditional measurement theory in favor of pragmatism for software engineering~\cite{Briand:1996:Application}.
The effect of this rejection is still felt today, with the topic of construct validity being remarkably absent in many software engineering papers, potentially leading to wrong conclusions~\cite{Ralph:2018:Construct}.
We therefore argue that, rather than rejecting the approaches from other fields, the community should embrace them.
For software metrics, methods and theories applied in psychology regarding construct validity could be particularly important in establishing empirical standards.

Notably, since the first advocacy for the evaluation of construct validity by Cronbach and Meehl in 1955~\cite{Cronbach:1955:Construct}, psychology has noticed significant improvements in clinical assessment through critical evaluation of all aspects of the construct validity process~\cite{Smith:2005:Construct}.
Some even call it \enquote{one of the most important concepts in all of psychology}~\cite{Westen:2003:Quantifying}.
And while the object of measurement differs and the measured concepts are often more concrete, many of the methodological guidelines should be applied for both human- and software-focused constructs.
Of particular importance is an explicit definition of the measured concept, and applying standardized tests and metrics to measure that concept, regardless of what is measured.
Recent works defining guidelines on evaluating construct validity in software engineering can be seen as valuable steps toward supporting this effort~\cite{Sjoberg:2023:Improving, Ralph:2018:Construct}.

Furthermore, the method of meta-analysis and the challenges of synthesizing data from heterogeneous studies have been an important topic in modern medicine research~\cite{DerSimonian:2015:Metaanalysis}.
The introduction of the random-effects model for the meta-analysis of clinical trials was intended to account for and explain the heterogeneity of studies due to inter-study differences in the employed methods or patient characteristics~\cite{DerSimonian:1986:Metaanalysis}.
Since then, the use of meta-analysis has been standard procedure in medicine and public health research, leading to modern approaches such as meta-regression and extensions of the random-effects model for multivariate meta-analysis~\cite{DerSimonian:2015:Metaanalysis}.
In addition to sophisticated statistical methods of synthesis, medical research exemplifies the adherence to strict methodological guidelines such as those outlined in the CONSORT statement~\cite{Moher:2001:Use}, and standardized terminology and measures that are published in various reporting guidelines~\cite{Simera:2010:Catalogue}.
Moreover, medicine has a larger focus on replication studies, further expanding the pool of data for established methodologies.
In software engineering, the opposite is the case: replication studies and meta-analyses are exceedingly scarce~\cite{Carver:2014:Replications, Jaccheri:2021:Systematizing}, and while a few standards and guidelines for measurement and experiment design exist, they are not universally followed.

\section{Ways Forward}
\label{sec:discussion}

It is not a novel finding that constructs exist in SE research and that they are measured in different ways.
Indeed, our community seems even so aware of them that there are guidelines written for SE research to ensure construct validity~\cite{Ralph:2018:Construct,Sjoberg:2022:Construct}, which should lead to meaningful and valid operationalization.
Furthermore, we were not the first to carry out a secondary study and, in fact, there are now guidelines for selecting an appropriate type of secondary research in software engineering as well~\cite{Ralph:2022:Paving}.
Practically speaking, however, the devil is in the detail when conducting a secondary study on certain constructs.
For example, we experienced the aforementioned challenge of deciding which primary studies are even comparable with each other on a conceptual level.
Based on our experience and what we can learn from other research fields, we see two concrete action items for the way forward that can make SE research more effective and efficient in the future.

\subsection{Define, Model, and Discuss Constructs}

We need clarity about the construct under investigation already during the design and execution of a primary study. Treating this as an afterthought during the study reporting is \textit{not} an acceptable practice. When using established constructs from other fields, such as personality and intelligence, definitions can usually be found and cited. In some cases, these constructs are based not only on dictionary-like definitions, but also on entire models that semantically describe the construct. In the case of personality, e.g., one could refer to the Five-Factor Model~\cite{McCrae:2020:FiveFactor}, and for intelligence, one could build on Carroll's Three-Stratum theory~\cite{Carroll:2005:ThreeStratum}, to explore a research question. If two studies cited and adhered to the same definition or model, it should be straightforward to argue that they are conceptually comparable.

Now, in the fast-paced field of software engineering, we do not always have the luxury of relying on established and well-defined constructs. Sometimes, a construct in a research field may be well-defined, but it has a slightly different meaning in the SE context. For example, text comprehension is a slightly older construct than code comprehension, and the code comprehension research drew on some ideas from its ancestor during its inception~\cite{Pearson:2015:FiftyReading}. However, conceptually, according to the current state of research, we are well-advised to define code and text comprehension as distinct constructs~\cite{Busjahn:2015:EyeLinearOrder,Peitek:2020:ReadingOrder,McChesney:2019:EyeDyslexia}. So, what do we do when such a definition is lacking?

Our suggestion is to propose missing construct definitions ourselves. This can result from a community effort, such as a workshop where experts on a construct participate. It can also arise from a public debate, in which individual researchers or research groups engage in discourse and articulate their proposals in position papers. We have observed that there is insufficient exchange among SE researchers regarding the semantics of abstract concepts, leading to uncertainty in the design of studies. Especially when there are no standardized tests, designing a primary study becomes a challenge in justifying one's operationalization of a construct if the construct has not been defined beforehand. Motivated by this, \citet{Wyrich:2023:CCDefinition} recently proposed a definition and a conceptual model for the concept of \emph{source code comprehension}. Secondary studies could make it a specific inclusion criterion that a primary study must be anchored in such a conceptual model to ensure comparability of included primary studies. We hope for more initiatives like these and advocate for a space for construct discussion within the respective research communities.

\subsection{Standardize and Validate Instruments}

Definitions help articulate the intentions of researchers, and we would appreciate it if each primary study defines its constructs. However, from the perspective of primary researchers, it remains laborious to develop a test for the construct under investigation for each study and potentially have to validate it as well. Therefore, the presence of standardized and validated tests, as known in psychology, would be a relief for primary research in SE. Secondary studies, in turn, could make use of a specific standardized test in primary studies as a criterion for inclusion to ensure comparability between primary studies. In some cases, there may still be the issue of an implicit definition of the construct through the tasks of such a test, but at least all included primary studies intended to measure \emph{something} in the exact same way.

An example from SE is the work of \citet{Siegmund:2014:Measuring} on measuring \emph{programming experience}. They developed a model of programming experience, designed a questionnaire, and followed a psychometrics approach to validate the questionnaire with computer science students. The work represents, according to the authors themselves, just the beginning of a larger effort to create a reliable and reusable test instrument for measuring programming experience. When the original conference article won the most influential paper award (MIP) at ICPC 2022, in the MIP Talk~\cite{SiegmundMIP:2021:Youtube}, Siegmund drew a somewhat sobering conclusion, stating that the test has not been further developed by either the authors or the community. One reason mentioned by Siegmund was that the validation of a measurement instrument often takes several years, and this is rarely appreciated by appointment or PhD committees. Another reason for the slow progress is that useful datasets exist but are not accessible. While many studies already build on the work of \citet{Siegmund:2014:Measuring}, these studies do not publish their own research data that could be used for further validation of the test instrument~\cite{SiegmundMIP:2021:Youtube}.

To make standardized measurement instruments for frequently studied constructs in the SE context a reality, a paradigm shift is needed within our research community. And there are some reasons to be optimistic about our methodological future. For example, SIGSOFT promotes its open science policies, e.g., at the International Conference on Software Engineering (ICSE), and several SE conferences now have dedicated replication tracks. It is our hope that this increasing recognition of open data and replications will over time increase the availability of data for validating measurement instruments in SE. At the same time, SE researchers need to be educated about the importance of construct validity and the significant value of validated measurement instruments. While it is completely normal in other disciplines to spend one or even several PhD projects on the development of a reliable and valid measurement instrument, a similar project would still raise the eyebrows of many SE researchers. In peer review, there is currently often not even a minimum requirement for construct validation on the part of the study authors. Additionally, in the use of validated tests, we often observe researchers modifying them without being aware of the consequences for construct validity. This is an indicator that guidelines and introductory articles on the subject are needed~\cite{Ralph:2018:Construct,Graziotin:2021:Psychometrics}.

\section{Conclusion}
When comparing primary studies on abstract concepts, there is a risk of literally comparing apples to oranges. In our opinion, there is nothing wrong with a fresh fruit salad. However, in SE research, it is currently challenging to discern the variety of fruits mixed in the bowl. Often, definitions for constructs are lacking, different terms are used interchangeably, and for hardly any construct, there exists a validated, reusable measurement instrument.

We have reported on two instances where the current situation posed challenges for us. Deciding which primary studies to select for a secondary study currently is, on a conceptual level, akin to a matter of intuition and laborious interpretative work. Related work from other fields has shown us that this situation can be improved, and that SE research can evolve methodologically.

Our resulting two-step guide to success can be summarized as follows: 1. Discuss, define, and model constructs within the respective research community, 2. Standardize and validate the measurement of these constructs. The outcome should benefit everyone involved in SE research: authors of primary studies, researchers conducting secondary studies, reviewers of both types of papers, and of course readers of SE papers.

\bibliographystyle{ACM-Reference-Format}
\bibliography{references}

\end{document}